\documentclass[12pt]{revtex4-2} 

\usepackage{filecontents}

\usepackage{mathtools}
\usepackage{amsmath}  
\usepackage{amsfonts} 
\usepackage{graphicx} 
\usepackage{xcolor}

\begin{document}


\title{Remark on the Entropy Production \\ of Adaptive Run-and-Tumble Chemotaxis
}


\author{Minh D. N. Nguyen}
\thanks{These authors contributed equally to this work.}
\affiliation{Department of Automation, School of Electrical and Electronic Engineering, \\ Hanoi University of Science and Technology, Hanoi 100000, Vietnam.  }

\author{Phuc H. Pham}
\thanks{These authors contributed equally to this work.}
\affiliation{High School for Gifted Students, Hanoi University of Science, Vietnam National University, 182 Luong The Vinh Str, Hanoi 100000, Vietnam.}

\author{Khang V. Ngo}
\affiliation{Ecole Polytechnique, Route de Saclay, Palaiseau 91120, France.}

\author{Van H. Do}
\affiliation{Homer L. Dodge Department of Physics and Astronomy, University of Oklahoma, 440 W. Brooks St. Norman, OK 73019, USA.}

\author{Shengkai Li}
\email{shengkaili@princeton.edu}
\affiliation{Department of Physics, Princeton University, Princeton, NJ 08544, USA.}

\author{Trung V. Phan}
\email{tphan23@jhu.edu}
\affiliation{Department of Chemical and Biomolecular Engineering, \\ John Hopkins University, Baltimore, MD 21218, USA.}

\begin{abstract}
Chemotactic active particles, such as bacteria and cells, exhibit an adaptive run-and-tumble motion, giving rise to complex emergent behaviors in response to external chemical fields. This motion is generated by the conversion of internal chemical energy into self-propulsion, allowing each agent to sustain a steady-state far from thermal equilibrium and perform works. The rate of entropy production serves as an indicates of how extensive these agents operate away from thermal equilibrium, providing a measure for estimating maximum obtainable power. Here we present the general framework for calculating the entropy production rate created by such population of agents from the first principle, using the minimal model of bacterial adaptive chemotaxis, as they execute the most basic collective action -- the mass transport. 
\end{abstract}
 
\date{\today}

\maketitle 

\section{Introduction}

\ \

Active matter are systems and materials made of autonomous energy-convertible entities engaged in self-organized collective motions \cite{schweitzer2003brownian,ramaswamy2010mechanics,marchetti2013hydrodynamics}. These constituents can be diverse, ranging from biological organisms like bacteria or cells \cite{hallatschek2023proliferating}, to synthetic constructs such as colloidal particles \cite{onsager1949effects,walther2008janus}, robots \cite{savoie2019robot,wang2021emergent,wang2022robots}, or even micro-robots \cite{miskin2020electronically}. Active particles operate in states far from thermal equilibrium, maintaining motility and functionality through the utilization of external forces, such as chemical gradients \cite{keller1971model,roussos2011chemotaxis} and temperature disparities \cite{mori1995neural,ozkan2021collective}. This capability enables these particles to access and execute tasks in complex environments and intricate geometries \cite{morris2017bacterial,phan2020bacterial} down to micron scales, where traditional energy sources may be limited or impractical. Research in active matter holds immense promise for a wide range of applications, including nano/biotechnology \cite{santiago2018nanoscale,ghosh2020active}, medicine and health care \cite{ornes2017medical,bunea2019strategies}, the development of novel smart materials with unique properties \cite{needleman2017active,fruchart2023odd}, and energy harvesting \cite{di2010bacterial,vizsnyiczai2017light}. Understanding and harnessing the physical principles underlying different families of active matter is a key area of research with significant potential for scientific discoveries and innovative technologies.

\ \ 

Quantifying extractable energy from active matter is of critical importance for engineering applications, enabling performance assessments, guiding optimal system designs, and facilitating comparative analysis. The actual power generation value $\dot{E}$ depends on the chosen work-extraction techniques, but a proportional estimation can be derived from the entropy production rate $\dot{S}_i$, i.e. $\dot{E} \propto \dot{S}_i$ \cite{ro2022model}. We can intuitively understand how this relation emerges by noting that the entropy production rate is a measure of departure from thermal equilibrium. When a system is distant from equilibrium, it signifies the existence of gradients or imbalances in energy transport within the system. These energy fluxes are what allow for work extraction when we couple degrees of freedom in the system with an external device, as outlined in \cite{anand2023bacteria}.
 We want to gauge $\dot{E}$, we need to estimate $\dot{S}_i$. Given the typical complexity of interactions among active particles, e.g. the emergence of non-reciprocity \cite{fruchart2021non}, simulations seem to be the most practical approach \cite{argun2021simulation}. Still, while physics {\it in silico} can provide answers quickly, valuable insights might be overlooked \cite{kuorikoski2013simulation}. Theoretical investigation from first principles allows for a more analytical and deeper understanding of the mechanisms governing active matter behavior, ensuring that engineering decisions are not only grounded in basic principles but also generalizable and flexible across a broader range of uses and settings.

\ \  

The aim of this brief report is to demonstrate how the entropy production rate can be derived from the first principles, for the ``most elementary form of active matter in nature'': chemotactic bacteria \cite{berg1975chemotaxis,erban2004individual,mattingly2021escherichia}. The behavior of these bacteria can be conveniently controlled by manipulating visualizable chemical concentrations, such as aspartate chemoattractants \cite{phan2023direct}. It should be noted that, although there has been previous attempt in this direction, the proposed estimation relies on a generic power-law assumption, $\dot{S}_i \propto n^\gamma$ (where $n$ denotes bacterial density and $\gamma$ is a constant exponent within the range of 0 to 1), without explicit grounding in the underlying physics \cite{vzupanovic2010bacterial}. Here, we showcase a concrete roadmap from the adaptive run-and-tumble model of chemotaxis \cite{long2017feedback,frankel2014adaptability,tu2013quantitative} to the analytical expression of $\dot{S}_i$ \cite{anand2023bacteria}, during steady collective migration \cite{fu2018spatial,mattingly2022collective}, which is arguably the most fundamental emergent behavior of a biological population \cite{keegstra2022ecological}. We start by constructing the Fokker-Planck equation to describe the probability distribution of all available bacterial motion states. We then introduce approximations using observations from wild-type bacteria to render the analysis more tractable; this enables us to derive an analytical solution that corresponds to steady collective transportation, which violates time-reversal symmetry. Finally, through the calculation of the time-irreversibility metric $\sigma$, we can establish a lower-bound estimation for the rate of entropy production $\dot{S}_i$. For simplicity, we only consider one-dimensional transport, but we mention how calculations can be done similarly in physical three-dimensional space. We also relates our estimated $\dot{S}_i$ with macroscopic quantities i.e. effective diffusivity $\mu$ and chemotactic coefficient $\chi$ of the bacterial field-theoretic dynamics as formulated in the Patlak-Keller-Segel partial differential equation (PDE) \cite{keller1971model,brenner1998physical,phan2023direct}.

\ \ 

\section{The Basics of Adaptive Run-and-Tumble Chemotaxis}

\ \ 

For a mathematical description of how chemotactic agents response to external chemical fields, let us utilize the minimal model of bacterial adaptive chemotaxis \cite{long2017feedback,frankel2014adaptability,tu2013quantitative}. In this model, we can describe the internal state of each agent by a single scalar value $F$. Biochemically, this value $F(t)$ represents the free energy difference between the active and inactive states of the bacterial receptors, measured in units of $k_{\text{B}} T$ (where $k_{\text{B}}$ is the Boltzmann constant and $T$ is the temperature of the environment). It quantifies the energy barrier or cost associated with transitioning between these states, playing a crucial role in governing the activity of these cooperative receptor clusters. Physically, this value $F(t)$ control the transition rates of switching from run to tumble $\lambda_{\text{R}}$ and tumble to run $\lambda_{\text{T}}$:
\begin{equation}
\lambda_{\text{T}}(f) = \frac{r(f)}{t_{\text{S}}} \ \ , \ \ \lambda_{\text{R}}(f) = \frac{1-r(f)}{t_{\text{S}}} \ \ , \ \ r(f) =\frac{1}{1+\exp\left(-f\right)} \ \ ,
\label{internal_update}
\end{equation}
where $t_\text{S}$ the timescale of switching between running and tumbling. Here we introduce $f=HF$, denoted as the gained internal state, where $H$ represents the flagella motor gain. This gain factor is determined by the kinetics governing how the concentration of response regulator inside the cell, controlled by receptor activity, binds to the flagellar motor complex. Say, at time $t$, the bacteria is at the position $\vec{x} = \vec{x}(t)$, then the rule for the adaptation of $F(t)$ follows:
\begin{equation}
\frac{d}{dt} F(t) = \frac{F_0-F(t)}{t_{\text{M}}} + N \left[ \partial_t + \frac{d}{dt} \vec{x}(t). \vec{\nabla} \right] \Phi \left[ C\left(\vec{x},t\right)\right] \Big|_{\vec{x} = \vec{x}(t)}   \ \ ,
\end{equation}
where $C(\vec{x},t)$ is the attracting chemical concentration and $\Phi\left[C(\vec{x},t)\right]$ is the signal strength as perceived by the bacterial receptors at that concentration value. Let us explain this elaborated equation, which describes the internal state $F(t)$ rate of change from two distinct contributions.  The first term reflects an adaptation process with a memory timescale $t_{\text{M}}$, governing the relaxation of $F(t)$ back to its innate value $F_0$. The second term outlines the updating mechanism which depends directly on how the cell perceives changes in the local environment over time $d\Phi/dt$, involving a gain factor $N$. It is important to note that, the internal state as observed in nature can be much more complicated, such as for {\it E.coli} bacteria, their internal state is often specified by two scalar values \cite{erban2004individual,setayeshgar2005application,lee2023learning}. 

\ \ 

When the agents run, we assume they always swim at a constant velocity $v_0$; when bacteria tumble, they stop and reorient the swimming direction. We consider a constant perceived signal gradient, so that the mass transport of this population will have a constant collective drift-velocity:
\begin{equation}
\Phi\left[C(\vec{x},t) \right] = \frac{z}{L} + \text{const}  \ , \ z =  \vec{x}.\hat{z} \ \ \Longrightarrow \ \ \vec{\nabla} \Phi\left[C(\vec{x},t)\right] = \frac{\hat{z}}L \ \ ,
\label{gradient_setting}
\end{equation}
in which $L$ represents the characteristic length of the gradient. For linear-sensitive agents $\Phi(C) \propto C$, hence the chemical concentration is just that of a constant slope. For log-sensitive agents $\Phi(C)\propto \ln(C)$ such as {\it E.coli} bacteria when the chemical concentration is inside their natural dynamical range, this setting corresponds to an exponential ramp $C(\vec{x},t) \propto \exp(z/L)$ \cite{dufour2014limits,shimizu2010modular,frankel2014adaptability}.

\ \ 

\section{A Steady Collective Transport \label{steady_1D}}

\ \ 

\begin{figure*}[!htbp]
\includegraphics[width= 0.8\textwidth]{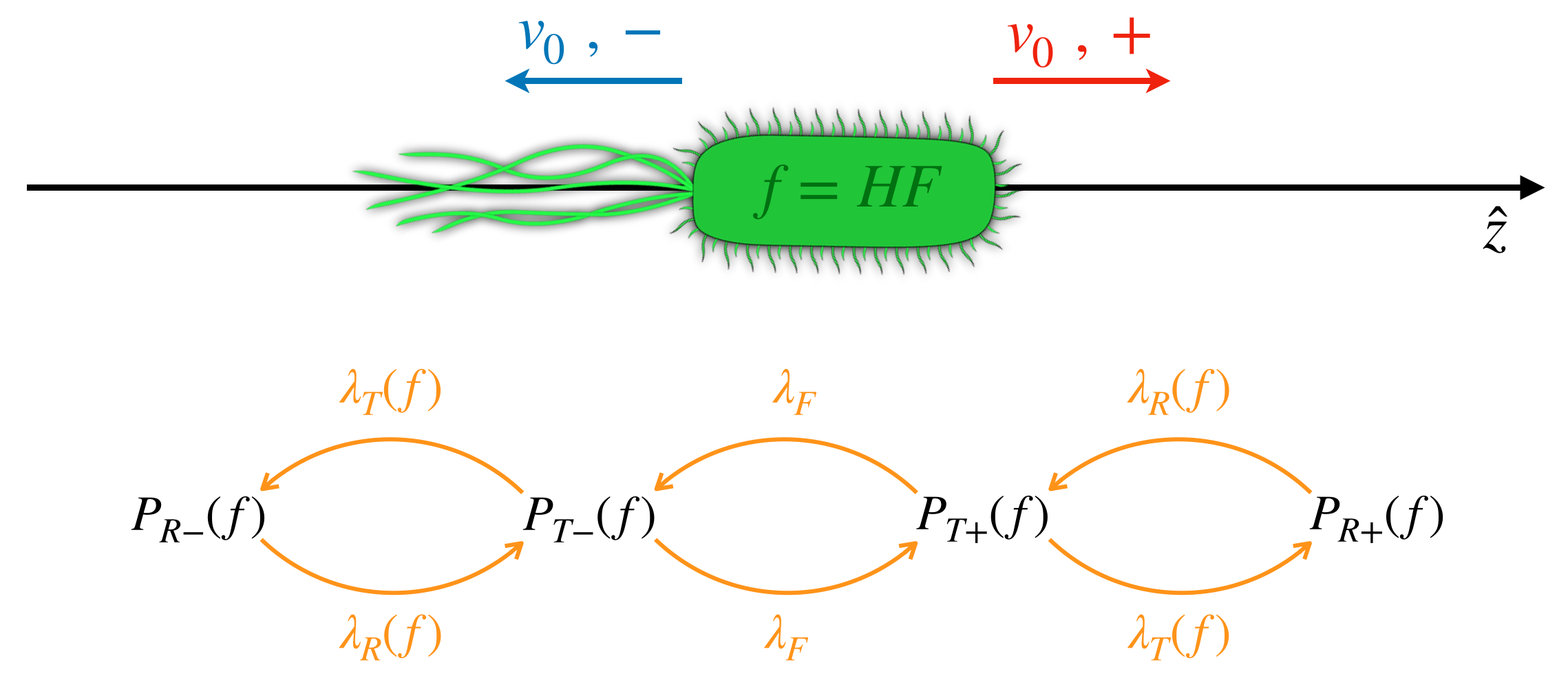}
\caption{Schematic of the one-dimensional adaptive chemotaxis.}
\label{fig01}
\end{figure*}

\ \ 

We can gain a lot of insights on the collective transport of adaptive chemotactic agents by consider the situation effectively in one-dimensional space. If the agents are only allowed to move in a single dimension, there are only four states of locomotion (see Fig. \ref{fig01}):
\begin{itemize}
    \item State $\text{R}-$: swimming in the $\hat{z}^-$ direction. The probability for a bacteria (at time $t$ located at position $x$, gained internal state $f$), to be in this state, is $P_{\text{R}-}(t,z,f)$.
    \item State $\text{T}-$: stopping while facing the $\hat{z}^-$ direction. The probability for a bacteria (at time $t$ located at position $x$, gained internal state $f$), to be in this state, is $P_{\text{T}-}(t,z,f)$.
    \item State $\text{T}+$: stopping while facing the $\hat{z}^+$ direction. The probability for a bacteria (at time $t$ located at position $x$, gained internal state $f$), to be in this state, is $P_{\text{T}+}(t,z,f)$.
    \item State $\text{R}+$: swimming in the $\hat{z}^+$ direction. The probability for a bacteria (at time $t$ located at position $x$, gained internal state $f$), to be in this state, is $P_{\text{R}+}(t,z,f)$.
\end{itemize}
Note that sum of all these probabilities should equal to unity:
\begin{equation}
1 = \sum^{\{ \text{R}-,\text{T}-,\text{T}+,\text{R}+\}}_{\alpha} \int df \int dz \int dt P_{\alpha} (t,z,f) \ \ .
\end{equation}

\ \ 

During stops, the facing direction can flip at a rate $\lambda_{\text{F}}=1/t_{\text{F}}$, where $t_{\text{F}}$ is the average flipping time. This value is an effective description, representing the directional persistency, which can be identified with the average time it takes for a $180^o$-angular change due to e.g. directional diffusion in physical space (as we will discuss in section \ref{probability_3D_diffusion_transport}). The set of Fokker-Planck's equations describing the evolution of these probabilities is given by:
\begin{equation}
\begin{split}
0 = \ &\partial_t P_{R\mp}\left(t,z,f\right) + \partial_z \left[ \mp v_0 P_{R\mp}\left(t,z,f\right)\right] + \partial_f \left[\left( \frac{f_0-f}{t_{\text{M}}} \mp \frac{NHv_0}{L} \right) P_{R\mp}\left(t,z,f\right)\right]
\\
+ & \left[ \lambda_{\text{R}}(f) P_{R\mp}\left(t,z,f\right) - \lambda_{\text{T}}(f) P_{T\mp}\left(t,z,f\right)  \right] \ \ ,
\\
0 = \ & \partial_t P_{T\mp}\left(t,z,f\right) + \partial_f \left[\left( \frac{f_0-f}{t_{\text{M}}} \right) P_{T\mp}\left(t,z,f\right)\right] 
\\ 
+ & \left[ \lambda_{\text{T}}(f) P_{T\mp}\left(t,z,f\right)  
- \lambda_{\text{R}}(f) P_{R\mp}\left(t,z,f\right)  + \lambda_{\text{F}} P_{T\mp}\left(t,z,f\right) - \lambda_{\text{F}} P_{T\pm}\left(t,z,f\right) \right] \ \ .
\end{split}
\label{all_prob_dyn}
\end{equation}
We show how these PDEs can be constructed in Appendix \ref{FokkerPlanck}. Define the ratios between time-scales:
\begin{equation}
    \tau_{\text{S}} = \frac{t_{\text{S}}}{t_{\text{M}}} \ \ , \ \ \tau_L = \frac{L/NHv_0}{t_{\text{M}}} \ \ , \ \  \tau_{\text{F}}= \frac{t_{\text{F}}}{t_{\text{M}}} \ \ .
    \label{timescale_list}
\end{equation}
When the switching rates between run and tumble is very fast $\tau_{\text{S}}\rightarrow 0$, we can assume to always have the equilibrium $P_{\text{R}\mp}(t,z,f) \lambda_{\text{R}}(f)\approx P_{\text{T}\mp}(t,z,f) \lambda_{\text{T}}(f)$. At the steady-state of collective transport, the probability density becomes stationary and homogeneous, thus we can drop the specifications $(t,x)$. If we  define $P_{\text{R}\mp}(f) + P_{\text{T}\mp}(f) = P_\mp(f)$, then we get $P_{\text{R}\mp}(f)=r(f) P_\mp(f)$, $P_{\text{T}\mp}(f)=\left[1-r(f)\right] P_\mp(f)$ and we can rewrite Eq. \eqref{all_prob_dyn} as:
\begin{equation}
0 \ = \ \partial_f \left\{ \left[ \left(f_0-f\right) \mp \frac{r(f)}{\tau_L} \right] P_{\mp}\left(f\right)\right\} + \left[ \frac{1-r(f)} {\tau_\text{F}} \right] \left[ P_\mp(f)-P_\pm(f) \right] \ \ .
\label{approx_prob_dyn}
\end{equation}
Note that, due to a finite swimming velocity, the available range for $f$ should be in between the values $f_\mp$ that satisfy:
\begin{equation}
f_\mp  = f_0 \mp \frac{r(f_\mp)}{\tau_L}  \ \ .
\label{f_range}
\end{equation}
As we can see from this expression, the ratio $1/\tau_L$ physically control the relative fluctuation of the internal state $f$ around its nature value $f_0$. When the population is climbing up a weak-gradient, $1/\tau_L \propto 1/L$ becomes small, therefore this range $[f_-,f_+]$ should be also be small.

\ \ 

The two ODEs in Eq. \eqref{approx_prob_dyn} can be solved analytically. From the integration $\Psi(f)$:
\begin{equation}
\Psi(f) =  \int^f df' \left\{ \frac{2\left[1-r(f')\right]} {\tau_\text{F}} \right\} \left\{\frac{\displaystyle (f'-f_0)}{\displaystyle \left[\frac{r(f')}{\tau_L} \right]^2 - (f'-f_0)^2} \right\}
 \ ,
\end{equation}
the probability distribution $P_\pm(f)$ can be written as:
\begin{equation}
P_\mp(f) \propto \left\{ \left[\frac{r(f)}{\tau_L}\right] \pm \left( f - f_0\right)\right\}^{-1} \exp\left[ - \Psi(f) \right] \ \ .
\label{prob_sol_1D}
\end{equation}
In the weak-gradient $1/L \rightarrow 0$ and fast-flipping direction limit, this becomes:
\begin{equation}
    P_\mp(f) \propto \left[ 1 \mp \left( \frac{f - f_0}{\alpha_L}\right) \right] \exp\left[-\frac{\alpha_{\text{F}}}2 \left( \frac{f - f_0}{\alpha_L}\right)^2 \right] \ ,
\label{approx_1D_prob_dis}
\end{equation}
where the coefficients $\alpha_L$ and $\alpha_{\text{F}}$ are given by:
\begin{equation}
\alpha_L = \frac{r(f_0)}{\tau_L} \rightarrow 0 \ \ , \ 
 \  
 \alpha_{\text{F}} =\frac{ 2\left[1-r(f_0)\right]}{\tau_{\text{F}}} \gg 1 \ .
\label{coeff_main}
\end{equation}
These are reasonable approximations, given that the length scale $L$ can be as long as we want, and for wild-type {\it E.coli} bacteria, $r(f_0)\sim 0.5$ to $0.8$, $t_\text{M} \sim 10$s to $30$s, and the directional reversal timescale $t_\text{F} \sim 1$s \cite{dufour2014limits,long2017feedback}, hence $\alpha_{\text{F}}$ can be as large as $30$. We give the derivation for this solution and its approximation in Appendix \ref{solving_approx_prob_dyn}. The peak positions of $P_\mp(f)$ distributions are located at $f_0 \mp \alpha_L/\alpha_{\text{F}}$, represents the expected bias for lower internal state value for bacteria swimming down-gradient and higher internal state value for bacteria swimming up-gradient. In other words, when bacteria swim up the gradient, they tend to tumble less.

\ \ 

\section{Entropy Production Rate}

\ \ 

\begin{figure*}[!htbp]
\includegraphics[width= \textwidth]{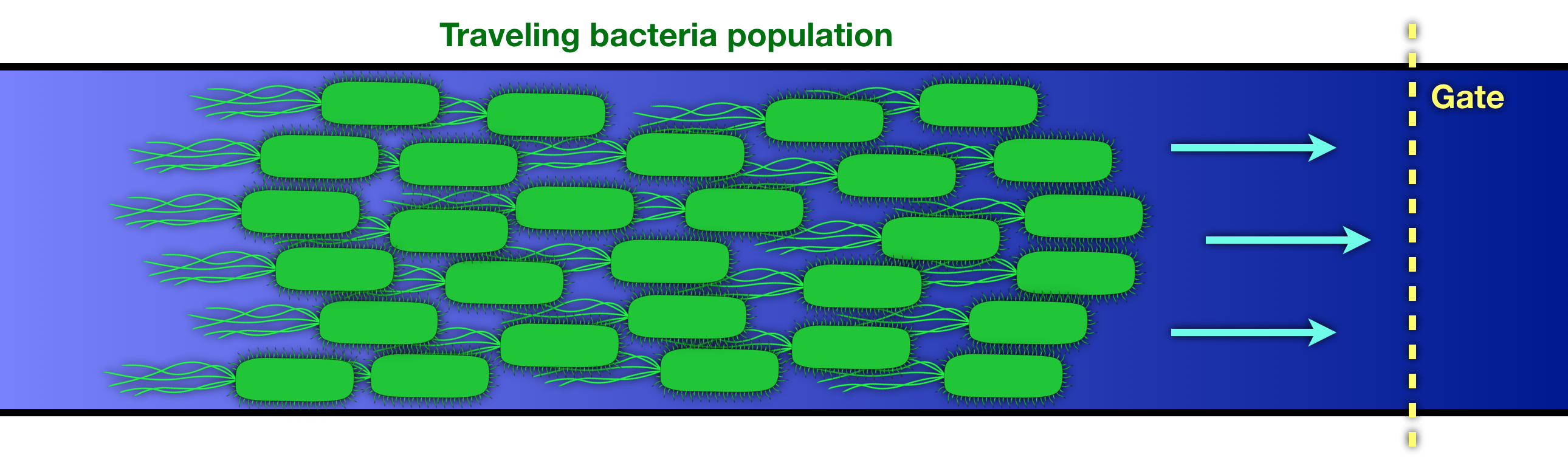}
\caption{The collective transport of bacteria. The entropy production rate $\dot{S}_i$ is quantified to be Eq. \eqref{entropy_production_rate_eq} at a gate where the energy can be harvested, e.g. as described in \cite{anand2023bacteria}.}
\label{fig02}
\end{figure*}

\ \ 

An engineer-definition of entropy production should be associated with how energy can be harvested out of the system. In \cite{anand2023bacteria}, a physical method is proposed for achieving this, involving a gate positioned perpendicular to the collective travel direction of the chemotactic agents, as illustrated in Fig. \ref{fig02}. In this proposal, the power generated by the population $\dot{E}$ is directly proportional to its entropy production rate $\dot{S}_i$, with the coefficient of proportionality depending on the specific harvesting technique employed. The entropy production rate can be estimated from the time-irreversibility metric $\sigma$ and the number of events that agents swim pass the gate per unit time $\Gamma$:
\begin{equation}
\dot{S}_i \geq k_{\text{B}} \Gamma \sigma \ .
\label{entropy_production_rate}
\end{equation}
The time-irreversibility metric $\sigma$ is calculated as the Kullback-Leibler divergence between the probability density functions of the time-forward and backward dynamics, quantifying how difference the collective behavior looks when playing the observations forward and backward in time \cite{ro2022model,anand2023bacteria,parrondo2009entropy,roldan2014irreversibility,roldan2010estimating,roldan2012entropy,gomez2008footprints,kawai2007dissipation,roldan2021quantifying,tan2021scale}.

\ \ 

Follow from Eq. \eqref{prob_sol_1D}, the probability for an agent swimming in the $\hat{z}^\mp$ direction is given by integrating all possible values of the internal state:
\begin{equation}
\mathcal{P}_\mp = \int^{f_+}_{f_-} df r(f)  P_\mp(f) \approx \frac12 r(f_0) \left( 1 \mp \frac{ \alpha_L}{\alpha_{\text{F}}} \right) \ \ ,
\label{1D_prob_forw_backw_eq}
\end{equation}
in which the approximation is calculated for the weak-gradient case. The time-irreversibility metric can be calculated by the Kullback-Leibler divergence between the probability density functions of chemotactic agents' trajectories in chronological and in reverse-chronological order:
\begin{equation}
\sigma = \mathcal{P}_+ \ln \frac{\mathcal{P}_+ }{\mathcal{P}_-} + \mathcal{P}_- \ln \frac{\mathcal{P}_- }{\mathcal{P}_+} \approx 2 r(f_0) \left( \frac{ \alpha_L}{\alpha_{\text{F}}} \right)^2 \ .
\label{time_irreversibility_metric}
\end{equation}
We derive these results in Appendix \ref{weak_grad}.

\ \ 

Consider a population of chemotactic agents with particle density $n$ per unit-length. At any point in space, the average rate of agents swimming pass it in the $\hat{z}^\mp$ direction is $\mathcal{P}_\mp n v_0$, thus the rate of gate-crossing events is $\Gamma = \left(\mathcal{P}_- + \mathcal{P}_+ \right)nv_0$. The entropy production rate of this chemotactic population as they pass through the gate shown in Fig. \ref{fig02} can be estimated using Eq. \eqref{entropy_production_rate}:
\begin{equation}
\dot{S}_i \geq k_{\text{B}} \left[ \left( \mathcal{P}_- + \mathcal{P}_+\right) nv_0\right] \left[ 2 r(f_0) \left( \frac{ \alpha_L}{\alpha_{\text{F}}} \right)^2 \right] \approx k_{\text{B}} \left\{ 2 n v_0 \left[ r(f_0) \frac{\alpha_L}{\alpha_\text{F}}  \right]^2 \right\} \ .
\label{entropy_production_rate_eq}
\end{equation}
Here, we found that the entropy production rate is proportional to the density of agents $n$, the inverse-squared of the directional flipping-rate $\alpha_{\text{F}}^{-2} \propto \tau_{\text{F}}^2 \propto \lambda_{\text{F}}^{-2}$, and the squared of the perceived signal gradient $\alpha_L^2 \propto (1/L)^2$.

\ \ 

Now let us connect our finding to the field-theoretic formulation. The diffusion coefficient $\mu$ and the chemotactic coefficient $\chi$ in the Patlak-Keller-Segel PDE \cite{keller1971model,brenner1998physical,phan2023direct} are related to the parameters of adaptive run-and-tumble chemotaxis:
\begin{equation}
\mu = \left\{ \frac{r^2(f_0)}{2\left[ 1-r(f_0) \right]} \right\} \frac{v_0^2}{\lambda_\text{F}} \ \ , \ \ \chi = \left( NH \right) \mu  \ \ ,
\end{equation}
which results we have derived in Appendix \ref{micro_to_macro_1D}. We can utilize this, along with Eq. \eqref{timescale_list} and Eq. \eqref{coeff_main} to rewrite the expression for the entropy production rate in Eq. \eqref{entropy_production_rate_eq}:
\begin{equation}
\dot{S}_i \geq \frac{2 k_{\text{B}} }{v_0} \left( \left\{ NH \frac{r^2(f_0)}{2[1-r(f_0)]} \right\} \frac{v_0^2}{\lambda_\text{F}}  \right)^2 \frac{n}{L^2} = \left( \frac{2k_{\text{B}} }{v_0} \right) n \left( \chi \nabla \Phi \right)^2 \ \ .
\label{PKS_entropy}
\end{equation}
We find that the estimated entropy production rate $\dot{S}_i$ is proportional to the density $n$ and the square of the perceived signal gradient $\nabla \Phi$.

\ \ 

It should be noted that there exists a simple argument for $\dot{S}_i \propto n \left( \nabla \Phi \right)^2$. Since the entropy of a classical system is an additive quantity of its constituents, it is expected that $\dot{S}_i \propto n$. If we further make a rather bold assumption that $\dot{S}_i$ is an analytical function in $\nabla \Phi$, a trick often used by Landau on emergent phenomena \cite{landau1937theory,landau1944problem}, then around $\nabla \Phi = 0$ we can approximate it with the Taylor expansion power-series:
\begin{equation}
\dot{S}_i \propto s_0 + s_1 \left( \nabla \Phi \right) + s_2 \left( \nabla \Phi \right)^2 + s_3 \left( \nabla \Phi \right)^3 + ... \ .
\label{EP_Taylor}
\end{equation}
When there is no gradient $\nabla \Phi = 0$ to create a preference direction in the environment, the flux of left-moving agents should equal to that of right-moving agents, and therefore the dynamics should have time-reversal symmetry $\dot{S}_i=0$ hence the constant term vanishes $s_0=0$. Also, from the non-negativity of the entropy production $\dot{S}_i \geq 0$ no matter the sign of $\nabla \Phi$, as followed from it being proportional to time-irrerversibility metric which is a Kullback-Leibler divergence \cite{ro2022model,anand2023bacteria}, the linear term must disappear $s_1=0$. So the first non-zero term should be at least of the second order, which means that in the limit $| \nabla \Phi | \rightarrow 0$, we must obtain $\dot{S}_i \propto \left( \nabla \Phi \right)^2$. Therefore, $\dot{S}_i \propto n \left( \nabla \Phi \right)^2$ is rather a reasonable guess.

\ \ 

\section{Considerations in Physical Three-Dimensional Space \label{probability_3D_diffusion_transport}}

For a realistic transport in physical space, which has three dimensions, the possible locomotion states of every agent are described by a continuous angular variable $\theta \in [0,\pi]$ which measures the angle between the facing direction and the uphill gradient direction $\hat{z}^+$, i.e. $P_\text{R}(\theta,f)$ for a run and $P_\text{T}(\theta,f)$ for a tumble at the steady-state, the combination $P(\theta,f)=P_\text{R}(\theta,f) + P_\text{T}(\theta,f)$ and the running probability $P_\text{R}\approx r(f) P(\theta,f)$. Due to a finite swimming velocity, the available range for $f$ should be in between the values $f_\mp$ that satisfy:
\begin{equation}
f_\mp  = f_0 \mp \frac{r(f_\mp)}{\sqrt{3} \tau_L}  \ \ .
\end{equation}
Moreover, instead of simply flip the facing direction as in one-dimensional space, the dynamics of reorientation depends on an angular diffusive timescale $\tau_\text{D}(f)$:
\begin{equation*}
\tau_\text{D} (f) = \frac{t_\text{D}(f)}{t_M} \ \ , \ \ t_\text{D}(f) = \Big( 2\left\{ r(f) D_{\text{R}} + \left[1-r(f) \right]D_{\text{T}} \right\} \Big)^{-1} \ \ ,
\end{equation*}
where $D_{\text{R}}$ is the angular diffusivity during a run and $D_{\text{T}}$ is the angular diffusivity during a tumble. This problem has been considered in \cite{long2017feedback}, required much more elaborated algebraic and analytical manipulations but the finding, while not exactly the same, has shared a lot of similar features to our effective one-dimensional toy-model, by replacing $\tau_{\text{D}}(f)$ with $\tau_{\text{F}}$. In other words, our model is a reasonable simplification of theirs.

\ \ 

The angular distribution $P(\theta,f)$ can be approximated by keeping the lowest-order projections on its spherical harmonics expansion:
\begin{equation}
P(\theta,f) \propto \left\{ \left[ \frac{r(f)}{\sqrt{3} \tau_L}\right] + \left(f-f_0\right) \cos\theta \right\} \left\{ \left[\frac{r(f)}{\sqrt{3} \tau_L} \right]^2 - (f-f_0)^2 \right\}^{-1} \exp\left[ - \Psi(f) \right] \ \ ,
\label{prob_sol_3D}
\end{equation}
in which we use the integration $\Psi(f)$:
\begin{equation}
\Psi(f) = \int^f df' \frac1{\tau_D(f')} \left\{\frac{\displaystyle (f'-f_0)}{\displaystyle \left[\frac{r(f')}{\sqrt{3} \tau_L} \right]^2 - (f'-f_0)^2} \right\} \ \ .
\end{equation}
In the weak-gradient $1/L \rightarrow 0$ and rapid reorientation limit, this becomes:
\begin{equation}
P(\theta,f) \propto \left[ 1 + \left( \frac{f - f_0}{\alpha_L}\right) \cos\theta \right] \exp\left[-\frac{\alpha_{\text{D}}}2 \left( \frac{f - f_0}{\alpha_L}\right)^2 \right] \ \ ,
\end{equation}
where the coefficients $\alpha_L$ and $\alpha_{\text{D}}$ are given by:
\begin{equation}
\alpha_L = \frac{r(f_0)}{\sqrt{3} \tau_L} \rightarrow 0 \ \ , \ 
 \  
 \alpha_{\text{D}} =\frac{1}{\tau_{\text{D}}(f_0)} \gg 1 \ .
\end{equation}
These are reasonable approximations, given that the length scale $L$ can be as long as we want, and for wild-type {\it E.coli} bacteria, $r(f_0)\sim 0.5$ to $0.8$, $t_\text{M} \sim 10$s to $30$s, $D_{\text{T}} \sim 2.2$rad$^2$/s to $4.4$rad$^2$/s and $\gg D_{\text{R}}$ \cite{saragosti2012modeling,dufour2014limits,long2017feedback}, hence it is possible for $\alpha_{\text{D}}$ to be more than an order of magnitude higher than unity. This is compatible with the coefficient of changing direction $\alpha_\text{F}$ in the effective one-dimensional space as we has studied in Section \ref{steady_1D}. The most probable $\max_f P(\theta,f)$ distribution are located at $f=f_0 + (\alpha_L/\alpha_{\text{D}})\cos\theta$, represents the directional bias for internal state value such that bacteria swimming up the gradient tumble less frequent.

\ \ 

From here on, we only sketch the steps. Follow from Eq. \eqref{prob_sol_1D}, the probability for an agent swimming in the $\theta$-direction is given by integrating all possible values of the internal state:
\begin{equation}
\mathcal{P}(\theta) = \int^{f_+}_{f_-} df r(f)  P(\theta,f) \ \ ,
\label{3D_prob_forw_backw_eq}
\end{equation}
in which the approximation is calculated for the weak-gradient case. The time-irreversibility metric can be calculated by the Kullback-Leibler divergence between the probability density functions of chemotactic agents' trajectories in chronological and in reverse-chronological order:
\begin{equation}
\sigma = \int^{\pi}_{0} \sin\theta d\theta \mathcal{P}(\theta) \ln \left[ \frac{\mathcal{P}(\theta)}{\mathcal{P}(\pi+\theta)} \right] \approx \left\{ \frac{r(f_0) \left[ 1 + r(f_0) \right]^2}{3\pi} \right\} \left( \frac{\alpha_L}{\alpha_\text{D}} \right)^2 \ \ .
\label{time_irreversibility_metric_3D}
\end{equation}
Consider the population density of $n$, corresponding to finding the average of $n$ agents per unit-volume. Through any surface perpendicular to $\hat{z}$, the average rate of agents swimming pass it per unit-area is given by:
\begin{equation}
\Gamma = nv_0 \int_0^{\pi} \sin\theta d\theta \mathcal{P}(\theta) \left| \cos\theta \right| \approx n v_0 \left[ \frac{r(f_0)}{4\pi} \right]  \ \ .
\end{equation}
The entropy production rate $\dot{S}_i$ of this chemotactic population per unit-area can be estimated using Eq. \eqref{entropy_production_rate}: 
\begin{equation}
\dot{S}_i \geq k_B \left( 3nv_0 \left\{ \frac{r(f_0) \left[ 1 + r(f_0) \right]}{6\pi} \left( \frac{\alpha_L}{\alpha_\text{D}} \right) \right\}^2 \right) \ .
\end{equation}
To relate this finding with the diffusivity $\mu$ and chemotactic coefficient $\chi$ in the Patlak-Keller-Segel field-theoretic model, we apply the results of Appendix \ref{micro_to_macro_3D} in the same way as how the answers of Appendix \ref{micro_to_macro_1D} can be used to derive Eq. \eqref{PKS_entropy}. We obtain a similar expression:
\begin{equation}
\chi = \left\{ NH \frac{r(f_0)\left[ 1 + r(f_0) \right]}{2 \sqrt{3} \pi} \right\} \mu \ , \ \dot{S}_i \geq \left( \frac{3 k_{\text{B}} }{v_0} \right) n \left( \chi \vec{\nabla} \Phi \right)^2 \ \ .
\label{PKS_entropy_3D}
\end{equation}
There is only a $1.5$ times difference between this answer of the full three-dimensional model and that of the simplified one-dimensional model, as given in Eq. \eqref{PKS_entropy}.

\ \ 

\begin{figure*}[!htbp]
\includegraphics[width=\textwidth]{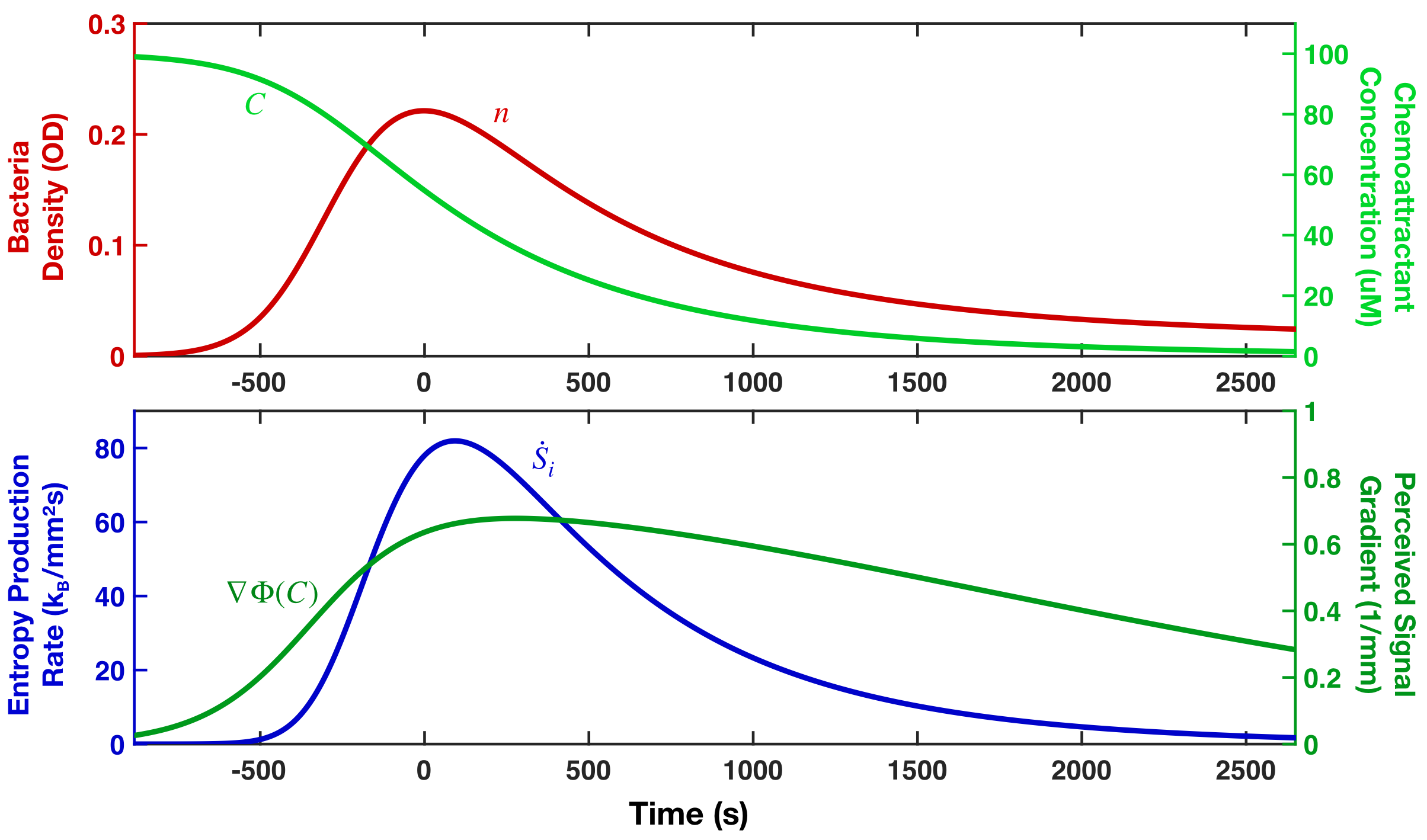}
\caption{A time-series of bacteria density $n(0,t)$ and chemoattractant concentration $C(0,t)$ measured at the gate $x=0$ (for energy harvesting i.e. Fig. \ref{fig02}) when a travelling solitary bacterial wave passes by, which dynamics is described by the set of PDEs in Appendix \ref{app:PKS_fig}. The temporal origin $t=0$ is chosen when the wave peak hits the gate. We can then get the perceived signal gradient $\nabla \Phi$ from the chemoattractant field profile $C(x,t)$, and estimate the entropy production rate $\dot{S}_i$ per unit area with Eq. \eqref{PKS_entropy_3D}. Note that $1$OD$=8 \times 10^8$cells/mL.}
\label{fig03}
\end{figure*}

The formulas presented by Eq. \eqref{PKS_entropy} and Eq. \eqref{PKS_entropy_3D} are extremely useful, since they can be used to estimate the entropy production rate and gauge the amount of extractable work from complex dynamical chemotactic systems. Let us demonstrate this. In Fig. \ref{fig03}, we give an example of how these equations can be applied to a self-generated (e.g. ``bootstrapped'' \cite{phan2021bootstrapped}) solitary wave of \textit{E.coli} bacteria, in which these cells follows the gradient of chemoattractants that they themselves create by consumption \cite{phan2023direct}. We find that the highest rate of bacterial entropy production occurs after the peak of the bacterial wave. From an informatic viewpoint, this suggests that the most information gained by the population comes from the pursuing bacteria. For real experiments, both bacteria density and the chemoattractant concentration can be measured directly in real time, allowing for a straightforward estimation of the entropy production rate even with a low-temporal data resolution \cite{phan2023direct}. In Appendix \ref{app:PKS_fig}, we discuss the underlying partial differential equations (PDEs) that govern the dynamics of bacteria-chemoattractant interaction depicted in this figure. We also provide the parameter values obtained through fitting with recent experiments \cite{phan2023direct}.

\ \ 

\section{Discussion}

\ \ 

We have demonstrated deriving the entropy production rate, from the first principles for a population of agents engaged in adaptive chemotaxis. This is done by utilizing a well-established relation between it and the time-irreversibility metric \cite{ro2022model,anand2023bacteria}, which in turn can be calculated from the probability distribution of all available locomotion states during a steady mass migration. This probability distribution is the stationary solution of the Fokker-Planck equation for an adaptive run-and-tumble random walk on a static environment of constant perceived signal gradient, leading to an unchanging average drift-velocity. Our bottom-up approach begins with the basics of motility then works upwards to obtain analytical results in simple regimes. While we focus on a one-dimensional space in our analysis, extending these calculations to higher dimensions is straightforward. We reframe our findings within the framework of the field-theoretic Patlak-Keller-Segel model, in which the entropy production rate should be assessed based on the effective diffusivity and the chemotactic coefficient of the whole populations, rather than relying on the length and time scales associated with bacterial trajectories.

\ \ 

We would like to emphasize that the approach present here is well-suited for characterizing systems where individual trajectories of active units can be easily identified. However, for systems that are better described by a field representation, a preferable approach would involve using a \textit{model-free} method. In this context, trajectories are not those of individual particles but rather time-series associated with the history (e.g. coarse-grained images or field-values) of local changes. A \textit{model-free} estimation for the entropy production rate, as previously developed \cite{ro2022model}, can then be defined from these time-series. In essence, rather than focusing on the directional probability of individual trajectories, understanding the temporal statistics of the field profile locally becomes crucial.

\ \ 

We have only begun to delve into the depths of a potentially rich and profound topic within theoretical biology. An important direction to explore in future works is to consider how the bacteria influence the environment. We should also take into account the effects of quorum sensing, which can be of extreme importance for many collective behavior of microbials \cite{surette1998quorum}. Our considerations for entropy production here works when the perceived signal gradient $\vec{\nabla} \Phi$ can be considered not only flat but also quasi-stationary, i.e. $| \partial_t \Phi | \ll v_0 | \vec{\nabla} \Phi |$. For example, collective transportation can emerge when bacteria chase after consumable resource, in which the flatness of $\vec{\nabla} \Phi$ emerges after the peak of the bacterial wave \cite{phan2023direct} and the quasi-stationary condition corresponds to $\left|\langle \vec{v} \rangle\right| \ll v_0$, as the drift velocity is much smaller than that of the swimming speed. This only holds true for small waves of bacteria migration, and beyond that is still an unexplored territory where we might also need to take into account hydrodynamic interactions between bacteria. Another direction worth investigating is for chemotaxis in natural complex geometries such as the fractal spaces of natural soils \cite{gimenez1997fractal}. Recent experimental studies using ``transparent soils'' have shown that bacterial motility changes drastically in porous environment \cite{bhattacharjee2019bacterial,bhattacharjee2021chemotactic}, and the field-theoretic Patlak-Keller-Segel formulation breaks down \cite{martinez2022morphological}. The effect of growth becomes important, thus instead of a trajectory-based method of quantifying entropy production as in here \cite{anand2023bacteria}, it is better to utilize a \textit{model-free} approach \cite{ro2022model}. For a bigger picture in a general dynamic environment, entropy serves as a conduit of information, allowing us to establish a measure of how the chemotactic population assimilates knowledge about its chemical surroundings. Moreover, considering entropy production enables us to pinpoint the specific locations where the intriguing collective behaviors, akin to the ``nucleation'' phenomena, originate \cite{ro2022model,anand2023bacteria}. Our small and simple exploration in this brief report has only offered a tantalizing glimpse into the coupling between biological systems and their environment, paving the way for further investigation and a deeper understanding of these complex phenomena from the first principles.

\ \ 

\section{Acknowledgement}

\ \ 

We thank Robert H. Austin for encouragement of this work. We also thank Stefano Martiniani and Satyam Anand for the many discussions about how bacterial entropy production rate can be estimated. S.L. was supported by the National Science Foundation, through the Center for the Physics of Biological Function (PHY-1734030) in Princeton University.

\ \ 

\appendix

\ \  

\section{Constructing the Fokker-Planck Equations \label{FokkerPlanck}}

\ \ 

Here we explain how the PDEs in Eq. \eqref{all_prob_dyn} are constructed. We start by noting that each chemotactic agent is specified by their position $z$ and their gained internal state $f$. The change in these values depends on their instantaneous velocity $v_i$ as follows:
\begin{equation}
\frac{d}{dt} z \Big|_i = v_i \ \ , \ \ \frac{d}{dt} f \Big|_i = \frac{f_0-f}{t_\text{M}} + \frac{NHv_i}{L}  \ \ .
\end{equation}
The later comes from Eq. \eqref{internal_update} and Eq. \eqref{gradient_setting}, using the definition $f(t)=HF(t)$. We use the label $i$ to denote the available motility states $\left\{ \text{R}-,\text{T}-,\text{T}+,\text{R}+\right\}$, then the drift terms for the probability $P_i (t,z,f)$ PDE is given by:
\begin{equation}
\partial_t P_i (t,z,f) = - \partial_z \left[ \frac{d}{dt}z \Big|_i P_i (t,z,f) \right] - \partial_f \left[ \frac{d}{dt}f \Big|_i P_i (t,z,f) \right] + \left( ... \right) \ \ .
\label{how_to_Fokk}
\end{equation}
For example, let us consider $i = \text{R}+$, thus $v_i = +v_0$. Then Eq. \eqref{how_to_Fokk} becomes:
\begin{equation}
\partial_t P_{\text{R}+} (t,z,f) = - \partial_z \left[ v_0 P_{\text{R}+} (t,z,f) \right] - \partial_f \left[ \left( \frac{f_0-f}{t_\text{M}} + \frac{NHv_0}{L} \right) P_{\text{R}+} (t,z,f) \right] + \left( ... \right) \ \ .   
\end{equation}
Do this for all other states, we can arrive at Eq. \eqref{all_prob_dyn}. The rest of the terms, which we refer to as $(...)$ in Eq. \eqref{how_to_Fokk}, are transition rates, as pictured in Fig. \ref{fig01}.

\ \ 

\section{Probabilistic Considerations in One-Dimensional Space \label{solving_approx_prob_dyn}}

\ \ 

\subsection{General Gradient $1/L$}

\ \ 
 
Let us look at the two ODEs as described in Eq. \eqref{approx_prob_dyn}, in details:
\begin{equation}
\begin{split}
        0 \ &= \ \partial_f \left\{ \left[ \left(f_0-f\right) - \frac{r(f)}{\tau_L} \right] P_{-}\left(f\right)\right\} + \left[ \frac{1-r(f)} {\tau_\text{F}} \right] \left[ P_-(f)-P_+(f) \right] \ \ ,  \\
        0 \ &= \ \partial_f \left\{ \left[ \left(f_0-f\right) + \frac{r(f)}{\tau_L} \right] P_{+}\left(f\right)\right\} + \left[ \frac{1-r(f)} {\tau_\text{F}} \right] \left[ P_+(f)-P_-(f) \right] \ \ .
    \end{split}
\end{equation}
By adding and subtracting these, we have an equivalent system:
\begin{equation}
    \begin{split}
        0 \ &= \ \partial_f \left\{ \left(f_0-f\right)\left[P_{-}\left(f\right)+P_{+}\left(f\right)\right]   + \frac{r(f)}{\tau_L}\left[-P_{-}\left(f\right)+P_{+}\left(f\right)\right]   \right\} \ \ ,
        \\
        0 \ &= \ \partial_f \left\{ -\left(f_0-f\right)\left[-P_{-}\left(f\right)+P_{+}\left(f\right)\right]   - \frac{r(f)}{\tau_L}\left[P_{-}\left(f\right)+P_{+}\left(f\right)\right]   \right\} 
        \\
        &- 2\left[ \frac{1-r(f)} {\tau_\text{F}} \right] \left[ -P_-(f)+P_+(f) \right] \ \ .
    \end{split}
\end{equation}
Define the sum $\mathbb{A}(f)= P_-(f) + P_+(f)$ and difference $\mathbb{S}(f)=- P_-(f) + P_+(f)$, which together can be used to obtain the probability density:
\begin{equation}
P_\mp(f)=\left[ \mathbb{A}(f)\mp\mathbb{S}(f)\right]/2 \ \ , 
\label{prob_for_sum_diff}
\end{equation}
then we get:
\begin{equation}
    \begin{split}
        0 \ &= \ \partial_f \left[- \left(f-f_0\right)\mathbb{A}(f)  + \frac{r(f)}{\tau_L}\mathbb{S}(f) \right] \ \ , 
        \\
        0 \ &= \ \partial_f \left\{(f-f_0) \mathbb{S}(f) -\frac{r(f)}{\tau_L} \mathbb{A}(f) \right\} - 2\left[ \frac{1-r(f)} {\tau_\text{F}} \right] \mathbb{S}(f) \ \ .
    \end{split}
\end{equation}
From the first equation we have the relationship:
\begin{equation}
\mathbb{A}(f) = \left[ \frac{r(f)}{\tau_L} \left(f-f_0 \right)^{-1} \right] \mathbb{S}(f)  \ .
\label{A_from_S}
\end{equation}
This can then be used in the second equation to arrive at:
\begin{equation}
0 \ = \ \partial_f \left[ \Xi^{-1}(f) \mathbb{S}(f) \right] + 2\left[ \frac{1-r(f)} {\tau_\text{F}} \right] \mathbb{S}(f) \ .
\label{diff_S}
\end{equation}
where the function $\Xi(f)$ is given by:
\begin{equation}
\Xi(f) = \frac{\displaystyle (f-f_0)}{\displaystyle \left[\frac{r(f)}{\tau_L} \right]^2 - (f-f_0)^2} \ .
\label{wanxue}
\end{equation}
We can now solve Eq. \eqref{diff_S} analytically, with an integration $\Psi(f)$ and a constant $\mathbb{C}$ yet to be determined:
\begin{equation}
\mathbb{S}(f) = \Xi(f) \exp\left[ - \Psi(f)  \right] \times  \mathbb{C} \ \ , \ \  \begin{dcases}
      &\Psi(f)\Big|_{f<f_0} =+\int^{f_0}_f df' \left\{ \frac{2\left[1-r(f')\right]} {\tau_\text{F}} \right\} \Xi(f')
      \\[10pt]
      & \Psi(f)\Big|_{f=f_0} = 0
      \\[10pt]
      &\Psi(f)\Big|_{f>f_0} =-\int^f_{f_0} df' \left\{ \frac{2\left[1-r(f')\right]} {\tau_\text{F}} \right\} \Xi(f')\end{dcases} \ . 
      \label{get_S_and_Psi}
\end{equation}

\ \ 

To determine $\mathbb{C}$, we make sure that the integral of $\mathbb{A}$ over all space is unity:
\begin{equation}
\begin{split}
    \int_{f_-}^{f_+} df'\mathbb{A}(f') &= \int_{f_-}^{f_+} df'\left[ \dfrac{r(f')}{\tau_L}(f'-f_0)^{-1}\right]\Xi(f') \exp\left[ -\Psi(f')  \right] \times  \mathbb{C}= 1 \\
    \Longrightarrow \mathbb{C}&=\left( \int_{f_-}^{f_+} df'\left\{ \frac{\displaystyle \left[\frac{r(f')}{\tau_L} \right]}{\displaystyle \left[\frac{r(f')}{\tau_L} \right]^2 - (f'-f_0)^2}\right\} \exp\left[-\Psi(f')\right] \right)^{-1} \ \ .
    \end{split}
    \label{get_C}
\end{equation}
Substitute into Eq. \eqref{prob_for_sum_diff} with our findings for $\mathbb{S}(f)$ as in Eq. \eqref{get_S_and_Psi} and $\mathbb{A}(f)$ as in Eq. \eqref{A_from_S}, we get the probability distribution: 
\begin{equation}
    P_\mp(f) = \left\{ \left[\frac{r(f)}{\tau_L}\right] \pm \left( f - f_0\right)\right\}^{-1} \exp\left[ - \Psi(f) \right] \ \times \ \frac{\mathbb{C}}2 \ \ ,
    \label{1D_prob_dis}
\end{equation}
which is the same as Eq. \eqref{prob_sol_1D}.

\ \ 

\subsection{Weak Gradient $1/L \rightarrow 0$ \label{weak_grad}}

\ \ 

Define the coefficients:
\begin{equation}
\alpha_L = \frac{r(f_0)}{\tau_L} \ \ , \ 
 \  
 \alpha_{\text{F}} =\frac{ 2\left[1-r(f_0)\right]}{\tau_{\text{F}}} \ .
 \label{alpha_coeffs}
\end{equation}
We can proceed further analytically by considering the weak-gradient limit $\tau_L \rightarrow \infty$ (as the characteristic length scale become extremely large $L \rightarrow \infty$), corresponding to $\alpha_L \rightarrow 0$. The available range as in Eq. \eqref{f_range} can be approximated as:
\begin{equation}
f_\mp = f_0 \mp \frac{r(f_\mp)}{\tau_L} \approx f_0 \mp \frac{r(f_0)}{\tau_L} = f_0 \mp \alpha_L \  ,
\end{equation}
where for the final manipulation we use $\displaystyle \frac{d}{df'} r(f') \Big|_{f'=f_0} = r(f_0) \left[ 1-r(f_0)\right]$ that follows from the functional form given in Eq. \eqref{internal_update}. Consider $\tau_{\text{F}} \ll 1$ which leads to $\alpha_{\text{F}} \gg 1$, then $P_\mp(f)$ is approximately a Gaussian function that dominates inside an even smaller region $|f-f_0| \sim \alpha_{\text{F}}^{-1/2}\alpha_L$ compared to the range $f \in [f_-,f_+]$ where $-f_-+f_+ \approx 2 \alpha_L $. We can then estimate $\Psi(f)$ from Eq. \eqref{get_S_and_Psi} as: 
\begin{equation}
\Psi(f) \approx \frac12 \alpha_{\text{F}} \alpha_L^{-2} (f-f_0)^2 + \varphi_3 (f-f_0)^3 + \varphi_4 (f-f_0)^4 + ...
\label{approx_the_Psi}
\end{equation}
where 
\begin{equation}
\varphi_3=\dfrac{1}{3}\alpha_{\text{F}} \alpha_L^{-2}[r(f_0)-2] \ \ , \ \ 
        \varphi_4=\dfrac{1}{8}\alpha_{\text{F}} \alpha_L^{-2}[2\alpha_L^{-2}+4-3r(f_0)] \ \ .
\end{equation}
Given that:
\begin{equation}
\begin{split}
& \left\{ \frac{\displaystyle \left[\frac{r(f')}{\tau_L} \right]}{\displaystyle \left[\frac{r(f')}{\tau_L} \right]^2 - (f'-f_0)^2}\right\} \approx 
\\
& \ \  
 \ \ \ \ \ \ \ \ \ \  \alpha_L^{-1}\left[1+C_1(f-f_0)+C_2(f-f_0)^2+3C_1C_2(f-f_0)^3+...\right] \ \ , \ \ 
 \end{split}
 \label{approx_the_ratio}
\end{equation}
in which the coefficients are:
\begin{equation}
C_1=r(f_0)-1 \ \ , \ \ 
 C_2=\alpha_L^{-2} \ \ ,
\end{equation}
we can now estimate the value $\mathbb{C}$ in Eq. \eqref{get_C}, using the approximations Eq. \eqref{approx_the_Psi} and Eq. \eqref{approx_the_ratio} to obtain a simple expression:
\begin{equation}
\begin{split}
\mathbb{C} &= \Big\{\alpha_L^{-1} \int_{-\infty}^{+\infty} \exp{\left[-\frac12 \alpha_{\text{F}} \alpha_L^{-2} (f-f_0)^2\right]} \times \\
 &\ \ \ \ \ \ \ \left[1+C_2(f-f_0)^2-(\varphi_4+\varphi_3 C_1)(f-f_0)^4\right]\Big\}^{-1} \\
 &\approx \left\{\sqrt{\dfrac{2\pi}{\alpha_\text{F}}}(1+\alpha_L^2\alpha_\text{F}^{-1}C_2 -3\alpha_L^4\alpha_\text{F}^{-2}(C_1\varphi_3+\varphi_4)\right\}^{-1} \approx \sqrt{\dfrac{\alpha_\text{F}   }{2\pi}}\dfrac{4\alpha_\text{F}}{4\alpha_\text{F} + 1} \ .
        \label{approx_the_C}    
\end{split}
\end{equation}

\ \ 

We also have the following expansion:
\begin{equation}
\begin{split}
&r(f) \left\{ \left[\frac{r(f)}{\tau_L}\right] \pm \left( f - f_0\right)\right\}^{-1} \approx  
\\
& \ \ \ \ r(f_0) \alpha_L^{-1} \left[ 1 \mp A_1 (f-f_0) + A_2 (f-f_0)^2 \mp A_3 (f-f_0)^3 + A_4(f-f_0)^4 ... \right] \ ,
\end{split}
\end{equation}
where 
\begin{equation}
    \begin{split}
        &A_1=\alpha_L^{-1}\\
&A_2=\alpha_L^{-1}\{\alpha_L^{-1}\pm [1-r(f_0)]\}\\
&A_3=\alpha_L^{-1}\left\{\alpha_L^{-2}+(\frac12 \pm 2\alpha_L^{-1})[1-r(f_0)]\right\}\\
&A_4=\alpha_L^{-1}\left[\alpha_L^{-3}\pm 3\alpha_L^{-2}+(2\mp3\tau_L)\alpha_L^{-1}+ (\tau_L^2\mp\dfrac{\tau_L}{6})\alpha_L-3\tau_L\pm \frac16\right] \ \ .
\end{split}
\end{equation}
Together with Eq. \eqref{approx_the_Psi} and Eq. \eqref{approx_the_C}, we obtain the backward and forward swimming probabilities:
\begin{equation}
\begin{split}
\mathcal{P}_\mp & = \int^{f_+}_{f_-} df r(f) P_\mp (f)
\\
&\approx \frac12 \mathbb{C}r(f_0)\alpha_L^{-1}  \int^{+\infty}_{-\infty} df \exp \left[ -\frac12 \alpha_{\text{F}} \alpha_L^{-2} (f-f_0)^2 \right] \ \times \ 
\\
& \ \ \ \ \ \ \ \ \ \  
 \ \left[ 1 + A_2(f-f_0)^2 + (A_4 - \varphi_4 \pm A_1 \varphi_3) (f-f_0)^4 \right]\\
&=\dfrac{\mathbb{C}\sqrt{2\pi}}{2\alpha_\text{F}^{5/2}}r(f_0)\left[\alpha_\text{F}^2+A_2\alpha_\text{F}\alpha_L^2+3\alpha_L^4(A4\pm A_1 \varphi_3-\varphi_4)\right]\\
 &\approx  \frac12 r(f_0)\left( 1 \mp \alpha_\text{F}^{-1}\alpha_L\dfrac{1-\dfrac{9}{\alpha_\text{F}}}{1+\dfrac{1}{4\alpha_\text{F}}} \right) \approx \frac{1}{2}r(f_0)\left[1\mp\alpha_\text{F}^{-1}\alpha_L\right] \ .
\end{split}
\end{equation}
From this we can calculate the average drift speed of this chemotactic population:
\begin{equation}
\langle v\rangle = v_0(\mathcal{P}_+-\mathcal{P_-}) \approx v_0r(f_0)\dfrac{\alpha_L}{\alpha_\text{F}} \ ,
\label{drift_speed}
\end{equation}
and the time-irreversibility metric:
\begin{equation}
    \sigma = \mathcal{P}_+ \ln \left(\dfrac{\mathcal{P}_+}{\mathcal{P}_-}\right) + \mathcal{P}_- \ln\left(\dfrac{\mathcal{P}_-}{\mathcal{P}_+}\right)= (\mathcal{P}_+ - \mathcal{P}_-)\ln \left(\dfrac{\mathcal{P}_+}{\mathcal{P}_-}\right) \approx 2r(f_0)\left(\dfrac{\alpha_L}{\alpha_\text{F}}\right)^2 \ .
\end{equation}

\ \ 

\section{From Micro to Macro}

\ \ 

The standard macro-description for chemotactic population is given by the Patlak-Keller-Segel model, where the density field of agents $n(\vec{x},t)$ follows a partial differential equation (PDE) consisting of a diffusion term $\propto \mu$, a chemotactic term $\propto \chi$ and a logistic growth term $\propto \alpha$ with carrying capacity $n_c$ \cite{narla2021traveling}:
\begin{equation}
\partial_t n(\vec{x},t) = \vec{\nabla}^2 \left[ \mu n(\vec{x},t) \right] + \vec{\nabla} \left\{ n(\vec{x},t) \chi \vec{\nabla} \Phi \left[ C(\vec{x},t) \right] \right\} + \alpha \left( 1 - \frac{n(\vec{x},t)}{n_c} \right) n(\vec{x},t) \ .
\label{PKS}
\end{equation}
Given the timescale of bacteria collective dynamics, the growth term is usually negligible compard to the other terms \cite{morris2017bacterial,phan2020bacterial,phan2023direct}. In the weak-gradient limit $\left| \vec{\nabla} \Phi \right| = 1/L \rightarrow 0$, it is possible to express the diffusivity $\mu$ and the chemotactic coefficient $\chi$ with the micro parameters of adaptive run-and-tumble. 

\ \ 

If bacteria perform chemotaxis involving multiple chemicals, each with concentration profile $C_i(\vec{x},t)$,  corresponding to a perceived signal $\Phi_i(C_i)$ and a specific chemotactic coefficient $\chi_i$, then we can generalize this model by combining the bias terms:
\begin{equation}
\chi \vec{\nabla} \Phi \ \ \longrightarrow \ \ \sum_i \chi_i \vec{\nabla} \Phi_i \ .
\end{equation}
For example, the entropy production rate in Eq. \eqref{PKS_entropy_3D} can therefore be estimated with:
\begin{equation}
 \dot{S}_i \geq \left( \frac{3 k_{\text{B}} }{v_0} \right) n \left( \sum_i \chi_i \vec{\nabla} \Phi_i \right)^2 \ .
\end{equation}

\ \  

\subsection{One-Dimensional Space \label{micro_to_macro_1D}}

\ \

The average drift-speed of the population as follows from Eq. \eqref{PKS} is $\langle v \rangle = \chi \nabla \Phi = \chi/L$, and we can match it with the results from Eq. \eqref{drift_speed} to obtain the chemotactic coefficient as: 
\begin{equation}
\chi = v_0 r(f_0) \frac{\alpha_L}{\alpha_\text{F}} L = \left\{ NH \frac{r^2(f_0)}{2\left[1-r(f_0) \right]} \right\} \frac{v_0^2}{\lambda_\text{F}}\ ,
\end{equation}
in which to arrive at the final answer we have used Eq. \eqref{alpha_coeffs}.

\ \ 

For the diffusivity $\mu$ in the weak-gradient limit, we can estimate that value by considering the diffusion of a homogeneous population with the same natural internal state $f=f_0$. We follow the approach outlined in \cite{lovely1975statistical}, where we treat running and tumbling as two simultaneous, independent Poisson processes. Specifically, during running, there is no change in direction. However, tumbling, which occurs at the fraction  $\left[ 1 - r(f_0) \right]$ of the total time, can lead to a switch in the direction of movement at a rate of $\lambda_\text{F}$. Thus, the correlation timescale $\tau_{\text{cor}}$ is given by:
\begin{equation}
\tau_{\text{cor}}= \left\{ 2\left[ 1 - r(f_0) \right]\lambda_\text{F} \right\}^{-1} \ \ .
\end{equation}
Use the average swimming speed $v_{\text{avg}} = v_0 r(f_0)$, in one-dimensional space we obtain:
\begin{equation}
\mu \approx v_{\text{avg}}^2 \tau_{\text{cor}} = \left\{ \frac{r^2(f_0)}{2\left[ 1-r(f_0) \right]} \right\} \frac{v_0^2}{\lambda_\text{F}} \ \ .
\end{equation}

\ \ 

\subsection{Physical Three-Dimensional Space \label{micro_to_macro_3D}}

\ \ 

Similar to Eq. \eqref{drift_speed}, in the three-dimensional space, the drift-velocity is given by:
\begin{equation}
\left| \langle \vec{v} \rangle \right| = v_0 \int^\pi_0 \sin\theta d\theta \mathcal{P}(\theta) \cos\theta \approx v_0 \left\{ \frac{r(f_0) \left[ 1 + r(f_0) \right]}{6\pi} \right\} \frac{\alpha_L}{\alpha_\text{D}}  \ \ .
\end{equation}
We match this with the populational average drift-speed as follows from Eq. \eqref{PKS}, which is $\left|\langle \vec{v} \rangle \right| = \chi \nabla \Phi = \chi/L$, to obtain the chemotactic coefficient as $\chi = \left| \langle \vec{v} \rangle \right| L$:
\begin{equation}
\chi = v_0 \left\{ \frac{r(f_0) \left[ 1 + r(f_0) \right]}{6\pi} \right\} \frac{\alpha_L}{\alpha_\text{D}} L = \left\{ NH \frac{r^2(f_0) \left[ 1+r(f_0)\right]}{12\sqrt{3}\pi \left[ 1-r(f_0) \right]} \right\} \frac{v_0^2}{D_T}   
\end{equation}

\ \ 

In the weak-gradient limit, to estimate the diffusivity $\mu$, we consider a homogeneous population of approximately identical internal state $f=f_0$. For $D_{\text{T}} \gg D_{\text{R}}$, meaning the rotational diffusivity in tumbling is much larger than that in running, we can directly utilize the result from \cite{saragosti2012modeling}:
\begin{equation}
\mu \approx \left\{ \frac{r(f_0)}{6\left[ 1-r(f_0) \right] } \right\} \frac{v_0^2}{D_{\text{T}}} \ \ .
\end{equation}

\ \

\section{A Realistic Model \\ for Self-Generating Travelling Solitary Waves of \textit{E.coli} Bacteria \label{app:PKS_fig}}

\ \ 

Here we use a model that fit reasonable well with the observed field dynamics at low-cell densities \cite{phan2023direct}. The PDE for bacteria dynamics is the same as Eq. \eqref{PKS}, with $n_c$ taken to be very large ($>5$OD in actual experiments, which is an order of magnitude higher than the peak density in Fig. \ref{fig03}). There, we formulate our mathematical model in effective one-dimensional space and fit the bacteria diffusivity with $\mu \approx 400\mu$m$^2$/s, the chemotactic coefficients $\chi\approx3300\mu$m$^2$/s, the growth rate $\alpha \approx 0.4$/hr. For the relevant range of chemoattactant concentration using in those experiments, which is aspartate at max $100\mu$M, the bacteria is log-sensitive to it. The perceived signal is given by:
\begin{equation}
\Phi(C) = \ln \left( \frac{1+C/K_i}{1+C/K_a} \right) \ \xrightarrow{ \ K_i \ll C \ll K_a \ } \ \ln(C) + \text{const} \ ,
\end{equation}
where $K_i \approx 1\mu$M and $K_a$ is very large (about $>50$mM) \cite{neumann2010differences,yang2015relation,moore2023sensory}. The swimming speed of these bacteria is about $v_0 \approx 30\mu$m/s.

\ \ 

The PDE for how chemoattractant concentration changes with time has only two contributions to this rate, which are diffusion term and consumption term:
\begin{equation}
    \partial_t C(x,t) = \nabla^2 \left[ \mu_C C(x,t)
 \right] - \gamma_0 \left[ \frac{C(x,t)}{C(x,t)+C_0} \right] n(x,t) \ , 
 \end{equation}
 where we have the diffusivity of aspartate molecules to be $\mu_C = 800\mu$m$^2$/s, fit the maximum consumption rate $\gamma_0$ with $0.45\mu$M/ODs and the half-max concentration $C_0$ with $7.6\mu$M.

\ \ 

The MatLab \cite{MATLAB} code used to generate Fig. \ref{fig03} will be provided by the corresponding authors upon reasonable request. This wave is travelling at speed $c=2.3\mu$m/s.

\ \ 

\bibliography{main}
\bibliographystyle{apsrev4-2}

\end{document}